\documentclass[conference]{IEEEtran}
\IEEEoverridecommandlockouts
\usepackage{bbm}
\usepackage{graphicx}

\usepackage{xcolor}
\usepackage{float}   
\usepackage[normalem]{ulem}
\usepackage{dsfont}
\usepackage{textcomp}
\usepackage{stfloats}
\usepackage{url}
\usepackage{verbatim}
\usepackage{algorithmic}
\usepackage{array}
\usepackage{booktabs}
\usepackage{pifont}
\usepackage{comment}
\usepackage{algorithm}
\usepackage{algorithmic}
\usepackage{amsthm}
\usepackage{amsmath,amsfonts, amssymb}
\usepackage{bm}
\usepackage{graphicx}
\newcommand{\argmax}{\mathop{\rm arg~max}\limits}

\usepackage[font=small]{caption}
\usepackage[font=footnotesize]{subcaption}
\usepackage{tikz}
\usetikzlibrary{plotmarks,patterns,decorations.pathreplacing,backgrounds,calc,arrows,arrows.meta,spy,matrix}
\usepackage{tikzscale}

\usepackage{pgfplots}
\usepgfplotslibrary{colorbrewer, groupplots, patchplots}
\pgfplotsset{
    compat=newest,
    legend style={font=\footnotesize, fill opacity=0.7,  draw opacity=1, text opacity=1, draw=white!15!black, legend cell align=left, align=left}, 
    width=0.8\columnwidth, 
    scale only axis,
    height=4cm,
    yminorticks=false,
    xminorticks=false,
    label style={font=\small},
    title style={font=\small},
    tick align=outside,
    tick pos=left,
    tick style={color=black},
    tick label style={font=\footnotesize},
    grid style={line width=.1pt, draw=gray!20},
    major grid style={line width=.1pt,draw=gray!20},
    plot coordinates/math parser=false % not sure if useful
}
\newlength\figureheight
\newlength\figurewidth

\usepackage{multirow}
\usepackage{tablefootnote}
\usepackage{booktabs}
\usepackage{tabularx}

\def\BibTeX{{\rm B\kern-.05em{\sc i\kern-.025em b}\kern-.08em T\kern-.1667em\lower.7ex\hbox{E}\kern-.125emX}}
\begin{document}

\title{Low-Power and Accurate IoT Monitoring Under Radio Resource Constraint}

\author{Takaho Shimokasa$^{\dag}$, Hiroyuki Yomo$^{\dag}$, Federico Chiariotti$^{\ddag}$, Junya Shiraishi$^{*}$, and Petar Popovski$^{*}$\\
        $^{\dag}$ Graduate School of Science and Engineering, Kansai University, Japan \\
        $^{\ddag}$ University of Padova, Italy \\
        $^{*}$Aalborg University, Denmark
\thanks{This work was, in part, supported by JSPS KAKENHI under Grant JP22K04114 and by the Velux Foundation, Denmark, through the Villum Investigator Grant WATER, nr. 37793. The work of J. Shiraishi was supported by Horizon Europe Marie Sk{\l}odowska-Curie Action (MSCA) Postdoc Fellowships with grant No.~101151067.}
   }

\maketitle

\begin{abstract}
This paper investigates how to achieve both low-power operations of sensor nodes and accurate state estimation using Kalman filter for internet of things  (IoT) monitoring employing wireless sensor networks under radio resource constraint. We consider two policies used by the base station to collect observations from the sensor nodes: \emph{(i) an oblivious policy}, based on statistics of the observations, and  \emph{(ii) a decentralized policy}, based on autonomous decision of each sensor based on its instantaneous observation. This work introduces a wake-up receiver and wake-up signaling to both policies to improve the energy efficiency of the sensor nodes. The decentralized policy designed with random access prioritizes transmissions of instantaneous observations that are highly likely to contribute to the improvement of state estimation. Our numerical results show that the decentralized policy improves the accuracy of the estimation in comparison to the oblivious policy under the constraint on the radio resource and consumed energy when the correlation between the processes observed by the sensor nodes is low. We also clarify the degree of correlation in which the superiority of two policies changes. 
\end{abstract}

\section{Introduction}
Wireless sensor networks (WSNs) play an important role as an enabling technology for remote internet of things (IoT) monitoring, in which physical processes observed by sensors located in a distant place are monitored in real time~\cite{IoT}\cite{Sensors}. The key requirements for WSNs that support IoT monitoring are: \emph{(i)} low-power operations of sensor nodes; \emph{(ii)} efficient use of radio resource required for data collection; and \emph{(iii)} accurate state estimation at a sink node.  

In this work, we focus on a scenario of IoT monitoring in which a base station (BS), which takes on a role of a sink, attempts to accurately estimate the states of sensor observations (i.e., the readings of sensor nodes) based on the Kalman filter by using data periodically collected from sensor nodes under radio resource constraint. The radio resource constraint comes from the fact that the channel is shared with other types of traffic, e.g., generated by smartphones, self-driving vehicles, etc., which limits the available amount of resource for data collection for IoT monitoring. In this case, the BS needs to keep track of sensor observations based on a subset of data collected from a part of the sensor nodes. In a conventional study on a similar scenario, the optimal scheduling policy is investigated, in which the BS selects the best sensor node to transmit data in a data collection period in terms of expected accuracy of estimation~\cite{Ramakanth_Infocom}. The authors of \cite{Ramakanth_Infocom} propose two scheduling policies: maximum expected error (MEE) and maximum weighted age (MWA) policies. Both of these policies are \textit{oblivious}, that is, the BS schedules a sensor node based on statistical information on sensor observations. While the MEE policy selects a sensor node based on the expected error of state estimation, the MWA policy uses the age of information (AoI)~\cite{AoI} as a scheduling measure. Then, it has been shown that these two policies are close to optimal when the processes monitored at the sensors are correlated, where an observation collected from a sensor node in a scheduling instance can be exploited to reduce the estimation errors of the other sensor nodes. On the other hand, for the uncorrelated case, a \textit{decentralized} policy based on ALOHA-like strategy has been developed in \cite{Non-oblivious}, in which its superiority to oblivious policy has been confirmed. However, these conventional studies do not take into account an important metric of sensor networks, that is, energy consumption. Furthermore, the impact of the degree of correlation between sensor observations on the superiority of two different policies is not investigated. 

In this paper, in order to clarify the above issues, we design scheduling policies including a low-power operation of the sensor node. We introduce a wake-up receiver and wake-up signaling~\cite{Wakeup_Survey}\cite{IEICE-yomo} into our system setting, which contributes to the reduction of the standby energy of each sensor node. We design decentralized and oblivious policies including the required mechanisms of wake-up signaling and radio access. The decentralized policy is based on a distributed decision of each sensor, in which each sensor node autonomously decides the necessity to transmit its instantaneous observation based on the possible deviation from its estimation at the BS. Each sensor node, which decides to transmit its data, employs a random access protocol to send its packet. Through numerical evaluations considering the communication and energy cost, we investigate the accuracy of the estimation and energy consumption of different policies. We evaluate the estimation accuracy under the same constraint on the radio resource and consumed energy for different policies, and clarify their superiority against the degree of correlation between sensor observations.

\section{System Model and Objective}
%In this section, we first describe our system model, followed by conventional estimation and scheduling framework considered in \cite{Ramakanth_Infocom}. Then, we clarify the objective of this work. 

\subsection{System Model}
\begin{figure}[t]
	\centering
	\includegraphics[width=0.35\textwidth]{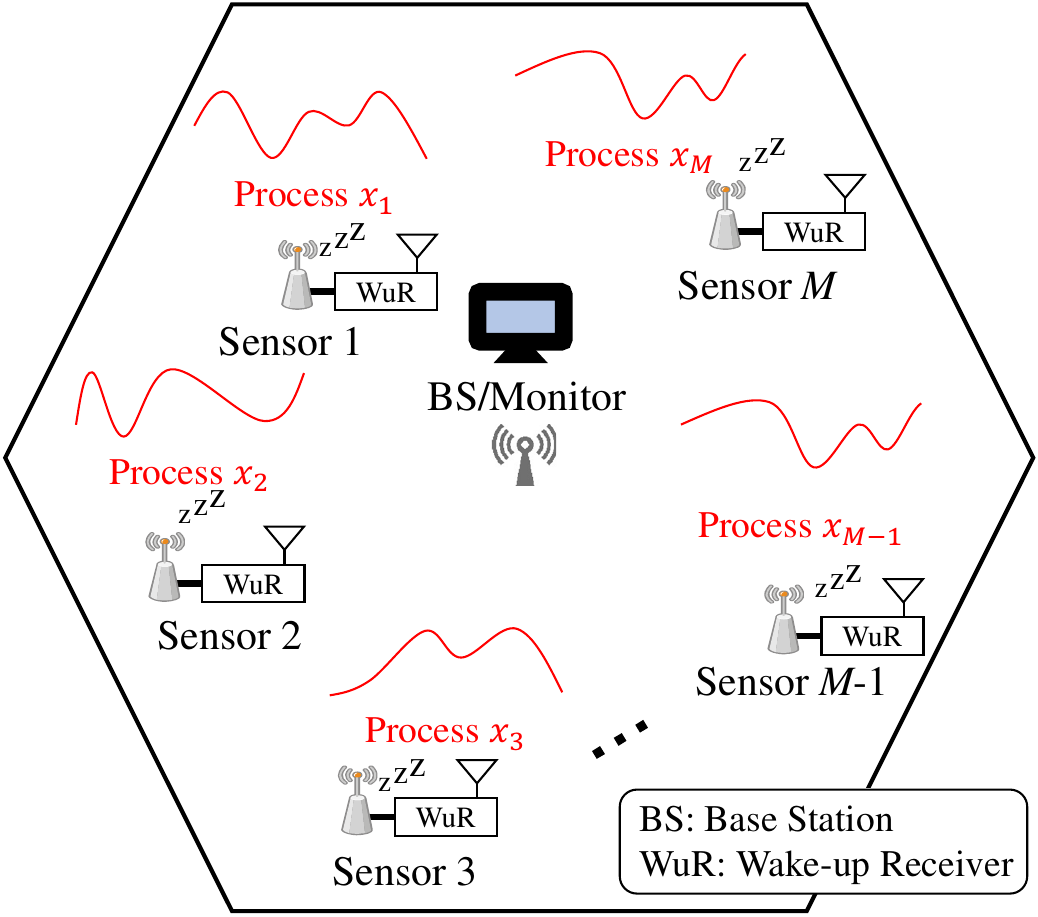}
	\caption{System Model.}
	\label{fig_systemmodel}
\end{figure}

\begin{figure}[t]
	\centering
	\includegraphics[width=0.45\textwidth]{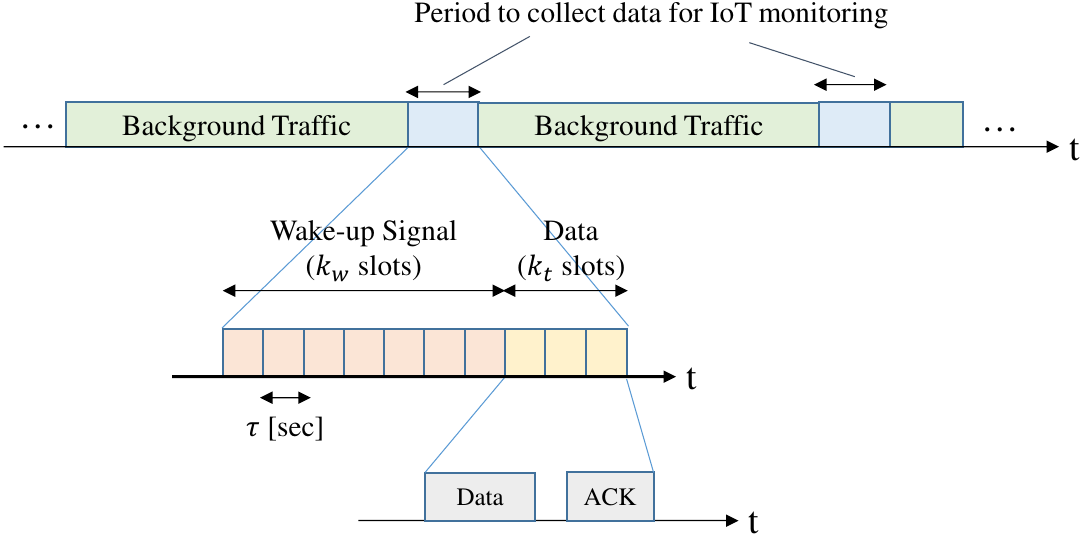}
	\caption{The slotted channel for the BS to collect data for IoT monitoring while sharing resource with the other background traffic.}
	\label{fig_channel}
\end{figure}

We consider a scenario in which $M$ sensor nodes are deployed within a cell managed by a single BS as shown in Fig.~\ref{fig_systemmodel}. The BS attempts to monitor the physical process observed by the sensor nodes. The sensor nodes are assumed to share radio resource with the other terminals, such as smartphones and vehicles; therefore, the BS is required to achieve accurate monitoring with limited resource allocated for data collections from sensor nodes. Since the power-saving of sensor nodes is of paramount importance, we assume that each sensor node employs a wake-up receiver~\cite{3GPP_Wakeup}: data communication module of each sensor node is activated by a wake-up signal transmitted by the BS. The wake-up receiver operating with ultra low-power consumption is always active while data communication module is only activated when polled by the BS through wake-up signaling, which allows for the energy-efficient operation of sensor nodes. Following the simplified model of 5G New Radio employed in \cite{SPAWC}, we assume a slotted channel with slot length of $\tau$ [s] to enable the BS to \textit{pull} the desired data from sensor nodes. This model assumes that $k_{w}$ slots are required for the BS to transmit a wake-up signal. This corresponds to, for example, on-off keying (OOK) based wake-up signal (WUS) transmitted over multiple orthogonal frequency division multiplexing (OFDM) symbols, which is considered in 3GPP Release 18~\cite{3GPP_R18}. Then, an activated node transmits data and receives the corresponding Acknowledgment (ACK) over the following $k_{t}$ slots. The frame structure is shown in Fig.~\ref{fig_channel}. For simplicity, we assume that the wake-up signaling is conducted error-free and that packets transmitted by the sensor nodes over the up-link are lost only due to collisions. Furthermore, we focus on a scenario where the IoT monitoring data is collected periodically by the BS while sharing the resource with the background traffic, as depicted in Fig.~\ref{fig_channel}.

We denote the state of physical process to be monitored at time $t$ as $\mathbf{x}(t)\in\mathbb{R}^M$. Following the simplification of discrete-time linear time-invariant (LTI) system model employed in \cite{Ramakanth_Infocom}, the processes are assumed to evolve as Gaussian random walks as
\begin{equation}
\mathbf{x}(t+1)=\mathbf{x}(t)+\mathbf{w}(t),
\end{equation}
where $\mathbf{w}(t)$ is a zero-mean multivariate normal random variable with a covariance matrix $\mathbf{Q}$ and assumed to be i.i.d. across time. Here, $\mathbf{Q}$ is expressed as
\begin{equation}
\arraycolsep=2pt
\mathbf{Q}=\begin{pmatrix} 
  q_{11} & \dots & q_{1j} & \dots & q_{1M} \\
  \vdots &       & \vdots &       & \vdots \\
  q_{i1} & \dots & q_{ij} & \dots & q_{iM} \\
  \vdots &       & \vdots &       & \vdots \\
  q_{M1} & \dots & q_{Mj} & \dots & q_{MM}
\end{pmatrix}  =
\begin{pmatrix} 
  q_{11} & \rho & \dots  & \rho \\
  \rho & q_{22} & \dots  & \rho \\
  \vdots & \vdots & \ddots & \vdots \\
  \rho & \rho & \dots  & q_{MM}
\end{pmatrix},
\label{covariance}
\end{equation}
where we assume the same degree of correlation denoted as $\rho$ between sensor observations.

We assume that the radio resource available for IoT monitoring is limited, which gives the constraint that only a part of the sensor nodes can be activated to transmit data during each period of data collection for IoT monitoring shown in Fig.~\ref{fig_channel}. Here, the goal of BS is to keep track of the overall state of $\mathbf{x}(t)$ with limited observation of the sensing data.

\subsection{Conventional Oblivious Scheduling for Kalman Filter Estimation}
In \cite{Ramakanth_Infocom}, the Kalman filter approach with the optimal scheduling policy is investigated for BS to track $\mathbf{x}(t)$ as accurately as possible. The work in \cite{Ramakanth_Infocom} considers that the scheduling instance (e.g., a period to collect data for IoT monitoring shown in Fig.~\ref{fig_channel}) is periodically repeated, where only one of the $M$ sensors can be requested to transmit state information. When the $j$-th sensor node is scheduled at time $t$, the BS receives a sample $y(t) \in \mathbb{R}$ denoted as
\begin{equation}
y(t)=\mathbf{c^\mathrm{T}}(t)\mathbf{x}_t,
\label{sample}
\end{equation}
where $\mathbf{c}(t)=\mathbf{e}_{j}= [0, \ldots,1,0, \ldots,0]^{T}$, and $\mathbf{e}_{j}$ is the $j$-th standard basis vector in $\mathbb{R}^M$. We denote a node scheduled at time $t$ as $\pi (t) \in \{1, ..., M\}$.

With the Kalman filter, the state estimation at time $t$ is recursively calculated using the estimate at time $t-1$ and the sample at time $t$ given in eq.~(\ref{sample}). Here, the estimation and error covariance matrix at time $t=0$, respectively denoted as
$\mathbf{\hat{x}}(0)$ and $\mathbf{P}(0)=\mathbb{E}[(\mathbf{x(0)-\mathbf{\hat{x}(0)}})(\mathbf{x(0)-\mathbf{\hat{x}(0)}})^\mathrm{T}]$, are assumed to be known. The Kalman filter algorithm consists of prediction and filtering steps as follows.
\paragraph{Prediction Step} Before using the current observation at time $t$, the prior estimate, $\mathbf{\hat{x}}^-(t)$, is updated by using the estimate at $t-1$, $\mathbf{\hat{x}}(t-1)$, as

\begin{equation}
\mathbf{\hat{x}}^- (t) = \mathbf{\hat{x}}(t-1).
\end{equation}
Furthermore, the prior covariance matrix, $\mathbf{P}^-(t)$, is calculated as

\begin{equation}
\mathbf{P}^-(t)=\mathbf{P}(t-1)+\mathbf{Q}.
\end{equation}

\paragraph{Filtering Step} In order to compute the estimate at time $t$ by using the prior estimate and current observation, the Kalman gain is calculated as follows:
\begin{equation}
\mathbf{g} (t) = \frac{\mathbf{P}^- (t) \mathbf{c} (t)}{ \mathbf{c}^\mathrm{T} (t)  \mathbf{P}^- (t)\mathbf{c} (t) }.
\label{KG}
\end{equation}
Then, the current estimate is updated as
\begin{equation}
\mathbf{\hat{x}} (t) = \mathbf{\hat{x}}^- (t)+\mathbf{g} (t) (y (t) - \mathbf{c^\mathrm{T}} (t) \mathbf{\hat{x}}^-(t) ).
\label{estimate}
\end{equation}
Finally, a posteriori covariance matrix, $\mathbf{P}(t)$, is updated as
\begin{equation}
\mathbf{P}(t)=(\mathbf{\it{I}}-\mathbf{g}(t)\mathbf{c^\mathrm{T}}(t))\mathbf{P}^-(t).
\end{equation}

In \cite{Ramakanth_Infocom}, two \emph{oblivious} policies in which the BS selects a sensor node based on statistical information on the accuracy of its estimation are considered for IoT monitoring employing the Kalman filter algorithm described above. The first policy is called the MEE policy, where a scheduled sensor node at time $t$, $\pi^{\mathrm{MEE}}(t)$, is selected as
\begin{equation}
\pi^{\mathrm{MEE}}(t) = \argmax_{i} \left( \frac{p_{ii}}{\sqrt{q_{ii}}} \right), 
\label{eq_MEE}
\end{equation}
where $p_{ii}$ is an element of the error covariance matrix $\mathbf{P}$.  
The second policy is called the MWA policy, where the following metric is used to select a sensor:
\begin{equation}
\pi^{\mathrm{MWA}}(t) = \argmax_{i} \left( \sqrt{q_{ii}}h_{i}(t-1) \right). 
\label{eq_MWA}
\end{equation}
Here $h_{i}(t)$ is AoI of sensor $i$, which is calculated as
\begin{eqnarray}
\it{h}_{i}(t)=
  \begin{cases}
    {h_i} (t-1)+1 & (\mathbf{c}(t)\ne \mathbf{e_i})\\
    0 & (otherwise).
  \end{cases}
  \label{AoI}
\end{eqnarray}

\subsection{Objective of This Work}
\label{sec_problem}
The above two policies, MEE and MWA, have been shown to be close to optimal when the processes monitored at the sensors are correlated. In this case, an observation collected from a sensor in a scheduling instance can be exploited to reduce the estimation errors of observations from the other correlated observations. Selecting a sensor based on statistical information as in MEE and MWA policies leads to a significant reduction of the expected error of estimations. On the other hand, when the processes are uncorrelated, exploitation of the other observation is not feasible. In this case, there is the potential for the BS to further improve the accuracy by collecting an observation that has the largest deviation from its estimation. However, the BS is oblivious to the instantaneous observation of each sensor upon scheduling. In order to realize data collections considering the instantaneous observation, a decentralized approach is required, in which each sensor autonomously decides whether it should transmit or not based on its possible deviation from the estimation at the BS. However, in this case, we also need a mechanism to avoid collisions as arbitrary sensors can contend to transmit packets containing their observations, which imposes communication cost to resolve contentions.  An ALOHA-like decentralized scheduling policy has been developed in~\cite{Non-oblivious}, and its superiority to oblivious policies has been confirmed for the uncorrelated setting.  

The above studies provide an important insight on the achieved accuracy of different scheduling policies, however, they lack the consideration on the energy efficiency that is an important metric of WSNs. 
Furthermore, the impact of the degree of correlation between sensor observations on the superiority of two different policies is not investigated. In this paper, we design scheduling policies including a wake-up signaling and radio access and evaluate the gain in terms of estimation accuracy, considering the cost of radio access and consumed energy for different degree of correlation.  

\section{Scheduling Policies}
In this section, we first design a decentralized policy including a wake-up signaling, sensor decision, and random access. Then, we show the oblivious policies fit into our framework.

\subsection{Decentralized Policy}
\begin{figure}[t]
	\centering
	\includegraphics[width=0.45\textwidth]{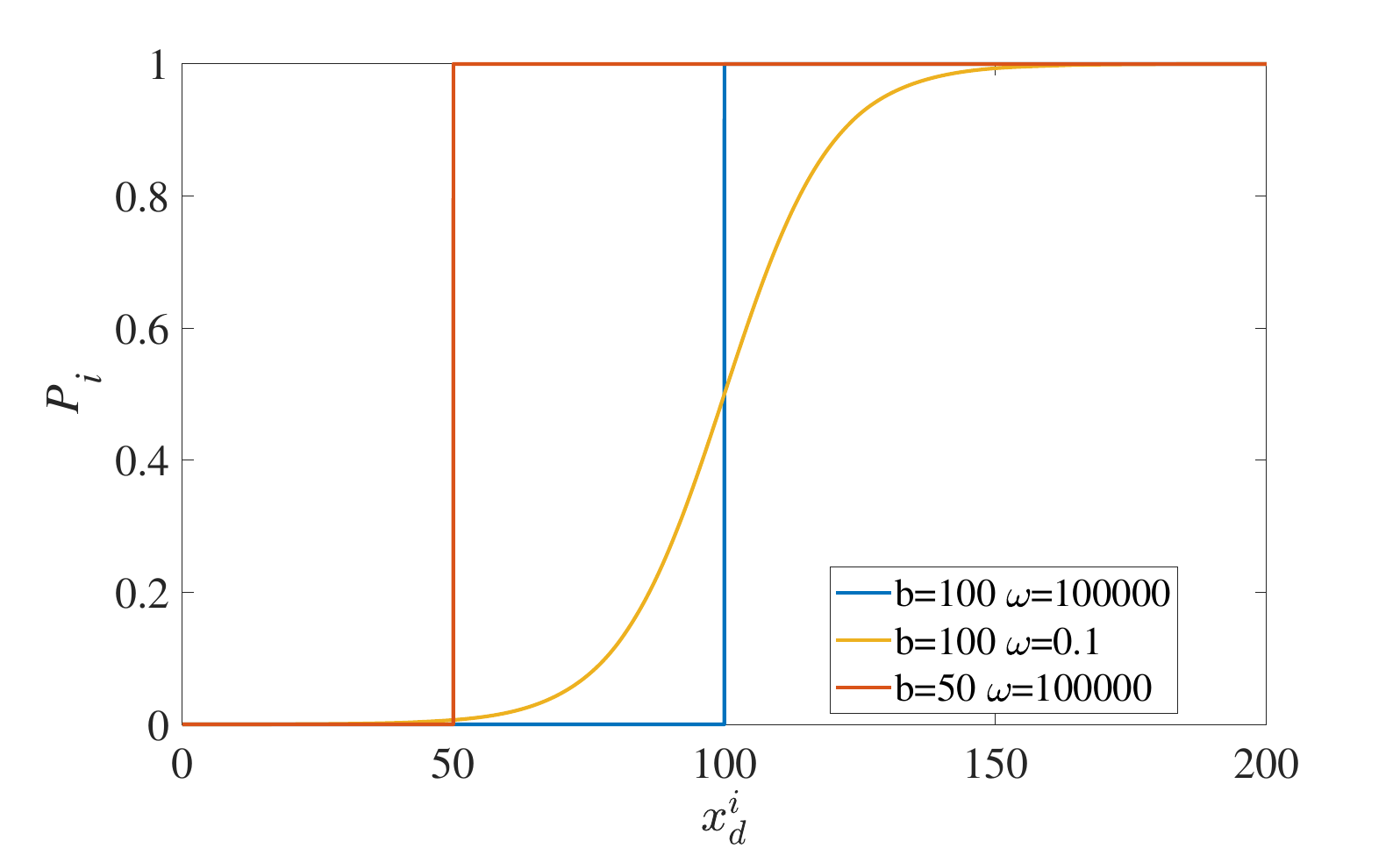}
	\caption{An example of Tx probability function with different values of $b$ and $\omega$.}
	\label{fig_tx_prob}
\end{figure}

We design a data collection scheme in which sensors that store observations with larger deviations from the estimation at the BS transmit their data with higher priority. We call this policy sensor-based decision (SBD). Let us denote the observation transmitted by the sensor $i$ in the last reporting instance as $x_{l}^{i}(t)$. Each sensor calculates the instantaneous error as the difference between its current observation, $x_{c}^{i}(t)$, and its last reported value, i.e., $e^{i}(t) = |x_{c}^{i}(t) - x_{l}^{i}(t)|$. Ideally, the BS should schedule a sensor node with the largest error normalized by $q_{ii}$, i.e., 

\begin{equation}
\pi^{\mathrm{SBD}}(t) = \argmax_{i} \left( \frac{e^{i}(t)}{\sqrt{q_{ii}}} \right). 
\end{equation}
However, the BS is oblivious to the instantaneous observation of $x_{c}^{i}(t)$. Therefore, our SBD policy employs a decentralized approach in which each sensor node decides its necessity to transmit data based on its instantaneous observation. To this end, we introduce a metric called normalized instantaneous error expressed as
\begin{equation}
x^i_d (t) =\frac{e^{i}(t)}{\sqrt{q_{ii}}}.
\label{eq_nie}
\end{equation}
Then, each sensor decides its necessity to transmit data based on $x^i_d (t)$ and a Tx probability function defined as
\begin{equation}
P_i=\frac{1-e^{-\omega x^i_d}}{1+e^{-\omega (x^i_d-b)}}.
  \label{eq_Tx_prob}
\end{equation}
Here, $b$ and $\omega$ are parameters to control the relationship between $P_{i}$ and $x^i_d$. An example of our Tx probability function with different values of $b$ and $\omega$ is shown in Fig.~\ref{fig_tx_prob}, which gives higher probability of $P_{i}$ for larger $x^i_d$. Each sensor node decides whether it should transmit data or not based on the derived $P_{i}$. This makes sensors with larger $x^i_d$ transmit their data with higher probability. As shown in Fig.~\ref{fig_tx_prob}, with larger value of $\omega$, Tx probability function behaves more like a threshold function, where, in this case, $b$ corresponds to a threshold. 

\begin{figure}[t]
	\centering
	\includegraphics[width=0.45\textwidth]{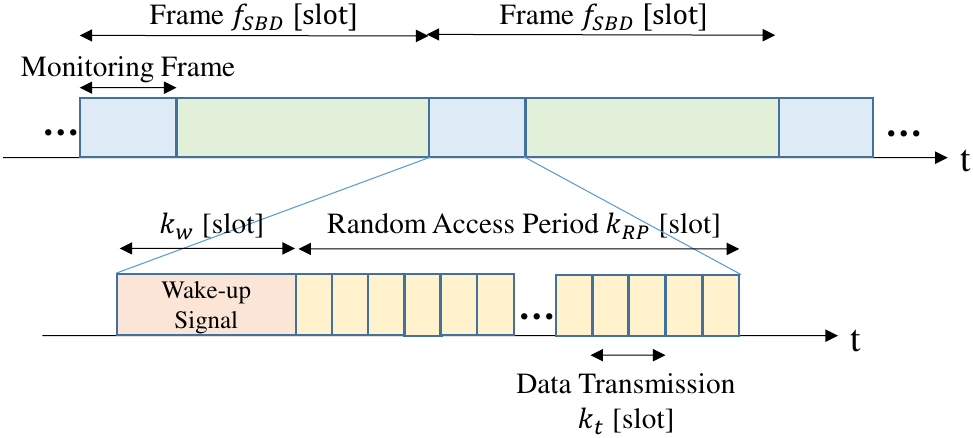}
	\caption{Frame structure of the SBD policy for data collection.}
	\label{fig_prop_frame}
\end{figure}

With the above SBD policy, it is possible that multiple nodes decide to transmit their data in response to a wake-up request from the BS. 
Therefore, we need a mechanism to resolve the contention among contending nodes. To this end, the SBD policy employs a frame structure shown in Fig.~\ref{fig_prop_frame}. In this frame structure, a period to collect data for IoT monitoring, which is denoted as \textit{monitoring frame}, is repeated every $f_{\mathrm{SBD}}$ [slot]. A monitoring frame is further divided into the period for wake-up signaling and data transmissions using random access. A wake-up signal is transmitted by the BS over $k_{w}$ [slot] at the beginning of a monitoring frame. This wake-up signal is used as a trigger for each sensor node to decide its necessity to transmit data based on eqs.~(\ref{eq_nie}) and (\ref{eq_Tx_prob}). When a wake-up receiver at each sensor node detects this wake-up signal, it decides whether to transmit data based on the derived $P_{i}$. If a sensor node decides to transmit data, it activates its main radio and randomly selects a slot to start transmission among $k_{\mathrm{RP}}$ [slot] during the random access period. After receiving an ACK from the BS, the sensor node transitions back to a sleep state in which its main radio is turned off and only the wake-up receiver is active. Note that, for simplicity, we assume that no retransmission is employed for the lost packet. After the successful reception of each sensor observation that corresponds to a sample given in eq.~(\ref{sample}), the Kalman filter algorithm is used to update the current estimate given in eq.~(\ref{estimate}) at the BS.

In the SBD policy, the number of slots within the random access period, $k_{\mathrm{RP}}$, plays an important role: $k_{\mathrm{RP}}$ should be increased to alleviate the negative impact of packet collisions, while its increase results in the longer monitoring frame, i.e., higher communication cost in terms of radio resources.   

\subsection{Oblivious Policy}
\begin{figure}[t]
	\centering
	\includegraphics[width=0.45\textwidth]{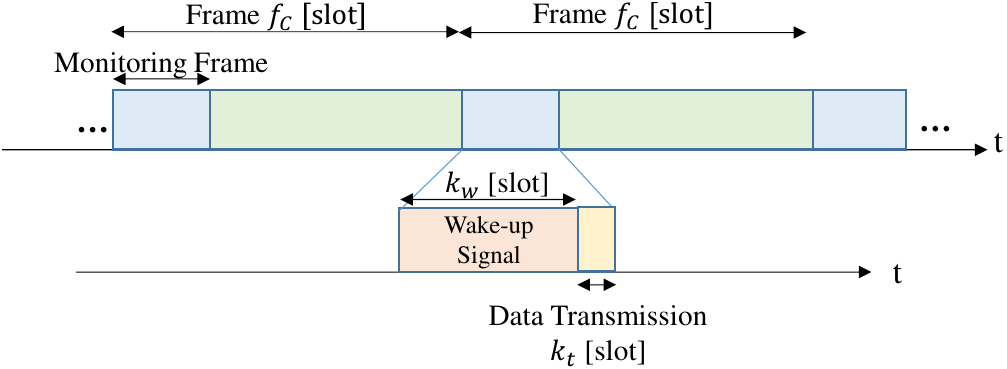}
	\caption{Frame structure of the oblivious MEE and MWA policies.}
	\label{fig_conv_frame}
\end{figure}

In order to fit the conventional oblivious policies into our framework, we apply the frame structure depicted in Fig.~\ref{fig_conv_frame}. Like the SBD policy, a monitoring frame is periodically repeated, consisting of slots for wake-up signaling and data transmissions. In the oblivious policies, the BS selects a single sensor node based on eq.~(\ref{eq_MEE}) or eq.~(\ref{eq_MWA}), and transmits a wake-up signal including an ID of the selected node, that is, employs a unicast wake-up~\cite{Shiraishi_TGCN}. Since only a single node is activated, the scheduled node can wake up and transmit its data in the following slot(s) without any protocol of contention resolution. 

\section{Numerical Results and Discussions}
\subsection{Simulation Model}
In this evaluation, we compare the estimation accuracy of the decentralized SBD policy and the oblivious MEE and MWA policies by setting their parameters in such a way that the fraction of time spent for collecting sensor observations (i.e., amount of radio resources) and the consumed energy are almost the same. The main parameters for the computer simulations are shown in Table~\ref{table_parameter}. Here, $f_{C}$ and $f_{\mathrm{SBD}}$ are set in such a way that the monitoring frame occupies $10\%$ of each frame, that is, $\it{f_C}$=10 ($\it{k_{w}}$+$\it{k_t}$)  [slot], and $\it{f_{\mathrm{SBD}}}$=10 ($\it{k_{w}}$+$\it{k_{\mathrm{RP}}}$) [slot]. The energy cost is defined as the total amount of energy consumed by the sensor nodes during the active state. We ignore the energy consumed by the wake-up receiver, as its value is much smaller than that of the main radio~\cite{Wakeup_Survey}. Thus, the energy cost [J] is calculated as
\begin{equation}
E =N_w\times{W_{a}}\times{\tau},
\label{eq_energy}
\end{equation}
where $N_w$ [slot] is the total number of slots in which sensor nodes are in active state and $W_{a}$ [W] is the power consumed by each sensor node in active state. 
The energy cost of the oblivious policies calculated based on the parameters given in Table~\ref{table_parameter} is 131.3 [mJ]. 
As for the performance metric of the accuracy of the estimation, we use the normalized error, which is the mean square error (MSE) normalized by the number of sensor nodes $M$. Here, MSE is the time average of the squared error between the true state of $\mathbf{x}(t)$ and the estimated state of $\mathbf{\hat{x}} (t)$ calculated in each slot. Note that, in this paper, we only present results when $\omega$ in Tx probability function is set to a large value, i.e., when the function behaves as a threshold function\footnote{This is because we have not observed significant impact of $\omega$ on the obtained results in our preliminary evaluations. The results will probably be different if we assume imperfect knowledge of the physical process at the BS, which will be kept for our future studies.}. 

\begin{table}[tb]
\begin{center}
\caption{Simulation Parameters.} 
\begin{tabular}{|c||c|}
\hline
%Parameter & Value\\ \hline \hline
Simulation Duration [slot]&$6.3\times10^5$\\ \hline
Total Duration of Monitoring Frame [slot]&$6.3\times10^4$\\ \hline
Slot Length $\tau$ [ms]&0.25~\cite{SPAWC} \\ \hline
Number of Sensor Nodes $M$ &[10, 150]\\ \hline
Duration for Wake-up Signaling $k_{w}$ [slot]&5 \\ \hline
Duration for Data Transmission $k_{t}$ [slot]&1 \\ \hline
Random Access Period $k_{\mathrm{RP}}$ [slot]& [15, 35] \\ \hline
Frame Length (Oblivious Policies) $f_C$ [slot]&$60$\\ \hline
Frame Length (SBD Policy) $f_{\mathrm{SBD}}$ [slot]&$ [200, 400] $\\ \hline
Power Consumption in Active State $W_{a}$ [mW]&50\\ \hline
$\omega$ in Tx Prob. Function &100000\\ \hline
$\it{b}$ in Tx Prob. Function & [5, 100] \\ \hline
Degree of Correlation $\rho$& [0, 0.5] \\ \hline
Variance $q_{ii}$& 1 ~\cite{Ramakanth_Infocom}\\ \hline
\end{tabular}
\label{table_parameter}
\end{center}
\end{table} 

\subsection{Results for the uncorrelated case}
\begin{figure}[t]
	\centering
	\includegraphics[width=0.5\textwidth]{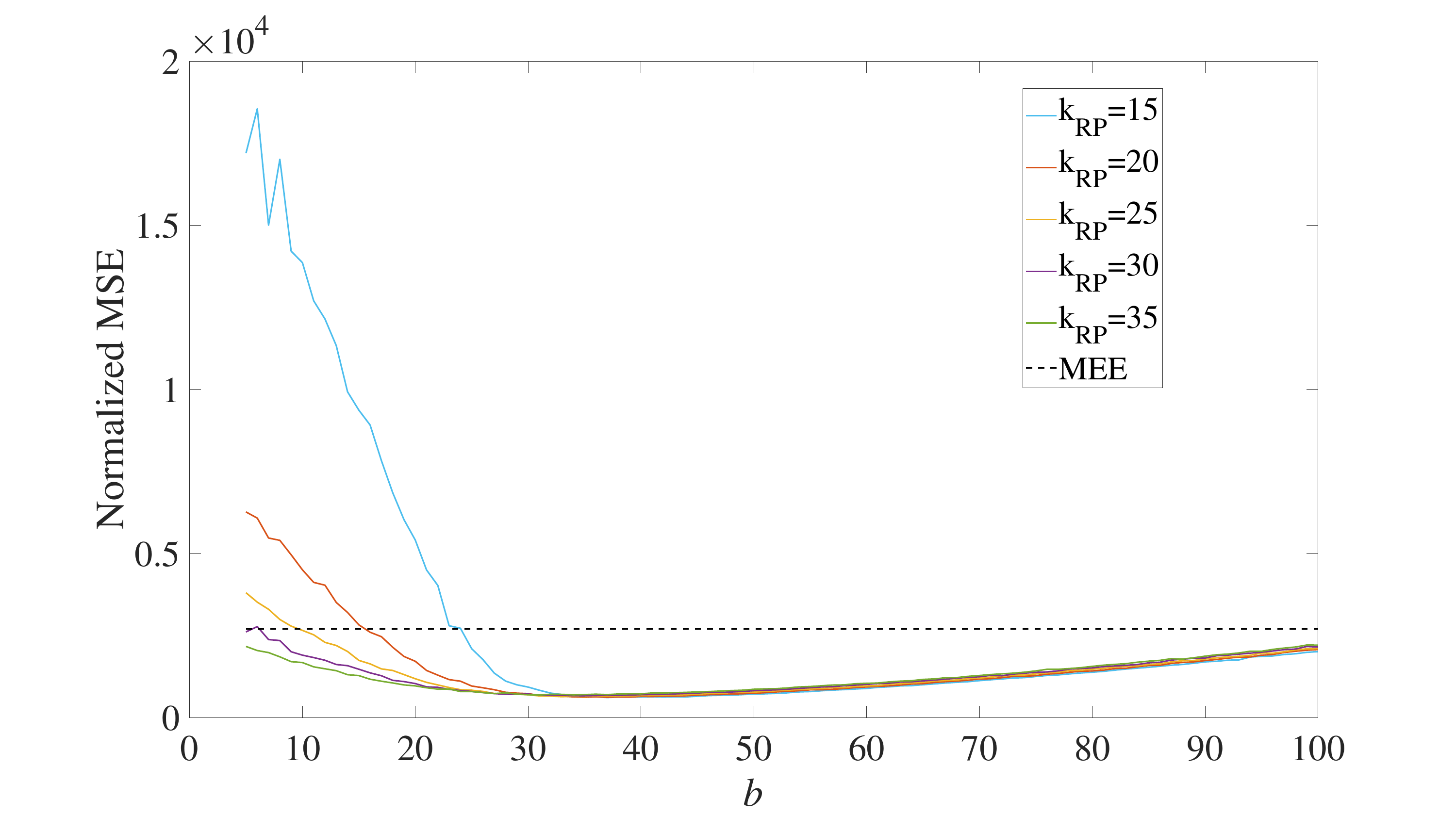}
	\caption{Normalized MSE of the SBD policy and the MEE policy against the parameter of threshold $b$.}
	\label{fig_error_b}
\end{figure}

\begin{figure}[t]
	\centering
	\includegraphics[width=0.5\textwidth]{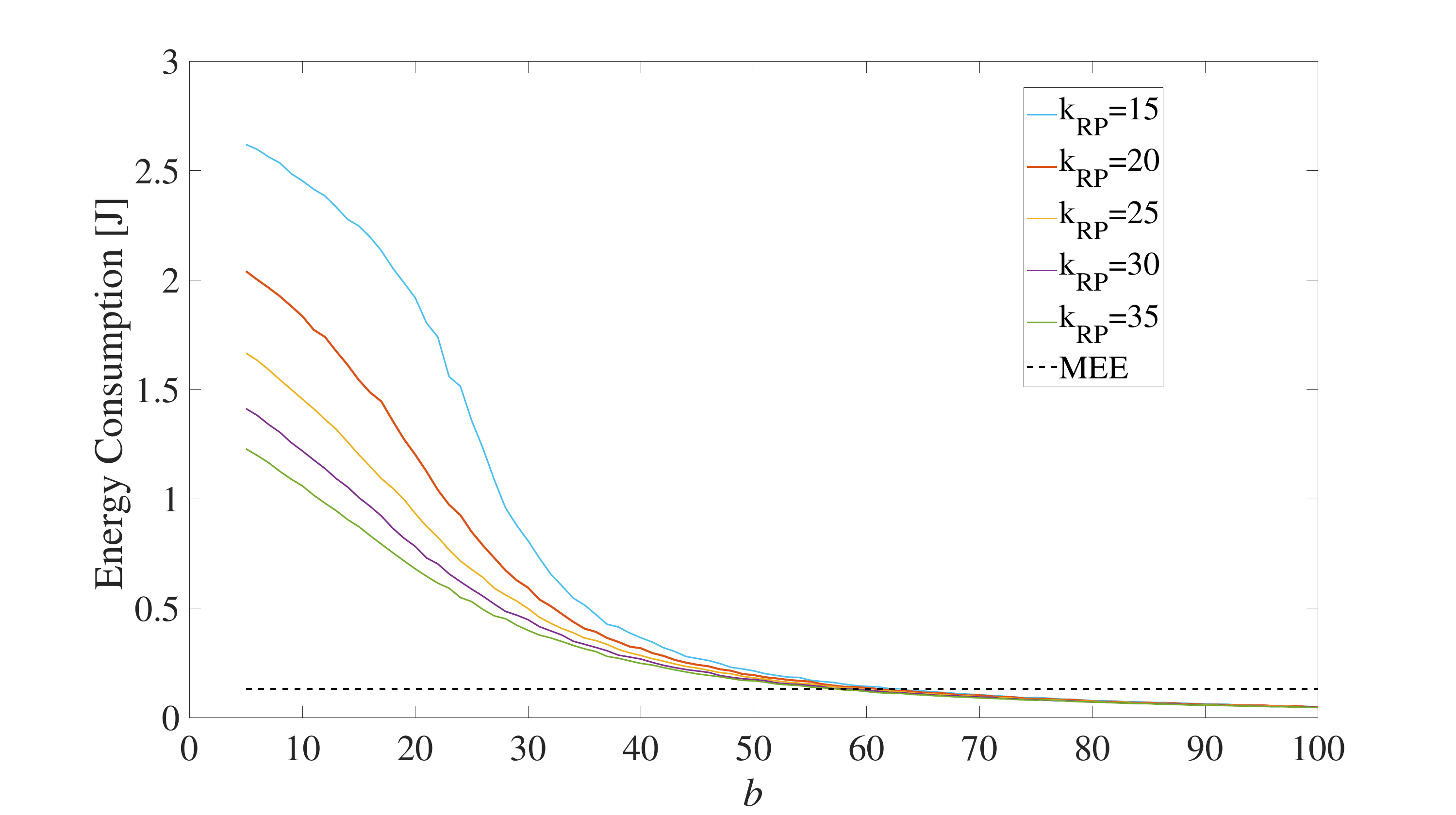}
	\caption{Energy Consumption of the SBD policy and the MEE policy against the parameter of threshold $b$.}
	\label{fig_energy_b}
\end{figure}

Here, we first evaluate the case where sensor observations are uncorrelated, i.e., the degree of correlation of $\rho$ = 0. Figs.~\ref{fig_error_b} and \ref{fig_energy_b} respectively show normalized MSE and energy cost of the SBD policy against the parameter of $b$ in the Tx probability function for different values of $k_{\mathrm{RP}}$. For reference, we also show the performance of the MEE policy in the same figures. First, from Fig.~\ref{fig_error_b}, we can see that the normalized MSE is larger for a smaller value of $b$. For small $b$, the number of nodes that store $x^{i}_{d}(t)$ larger than $b$ is large, causing severe congestion among the contending nodes. Therefore, many packets are lost due to collisions, deteriorating the accuracy of the estimation. By increasing $b$, the number of activated nodes decreases and the negative impact of the collision is alleviated, resulting in the decrease of the normalized MSE for $b < 35$. However, a further increase of $b$ overly reduces the number of sensor nodes to transmit data, again increasing the normalized MSE. Looking at the impact of $k_{\mathrm{RP}}$, we can see that the larger value of $k_{\mathrm{RP}}$ gives a smaller normalized error for $b <35$ since the larger number of slots prepared during the random access period contributes to the reduction of collision. However, for $b > 35$, larger $k_{\mathrm{RP}}$ results in worse performance since it leads to a longer frame duration of $f_{\mathrm{SBD}}$, which reduces the frequency of state update. Next, in Fig.~\ref{fig_energy_b}, it can be seen that the energy cost decreases monotonically as $b$ increases. This is simply because the number of sensor nodes that wake up for data transmissions is reduced with increasing $b$.

\begin{figure}[t]
	\centering
	\includegraphics[width=0.5\textwidth]{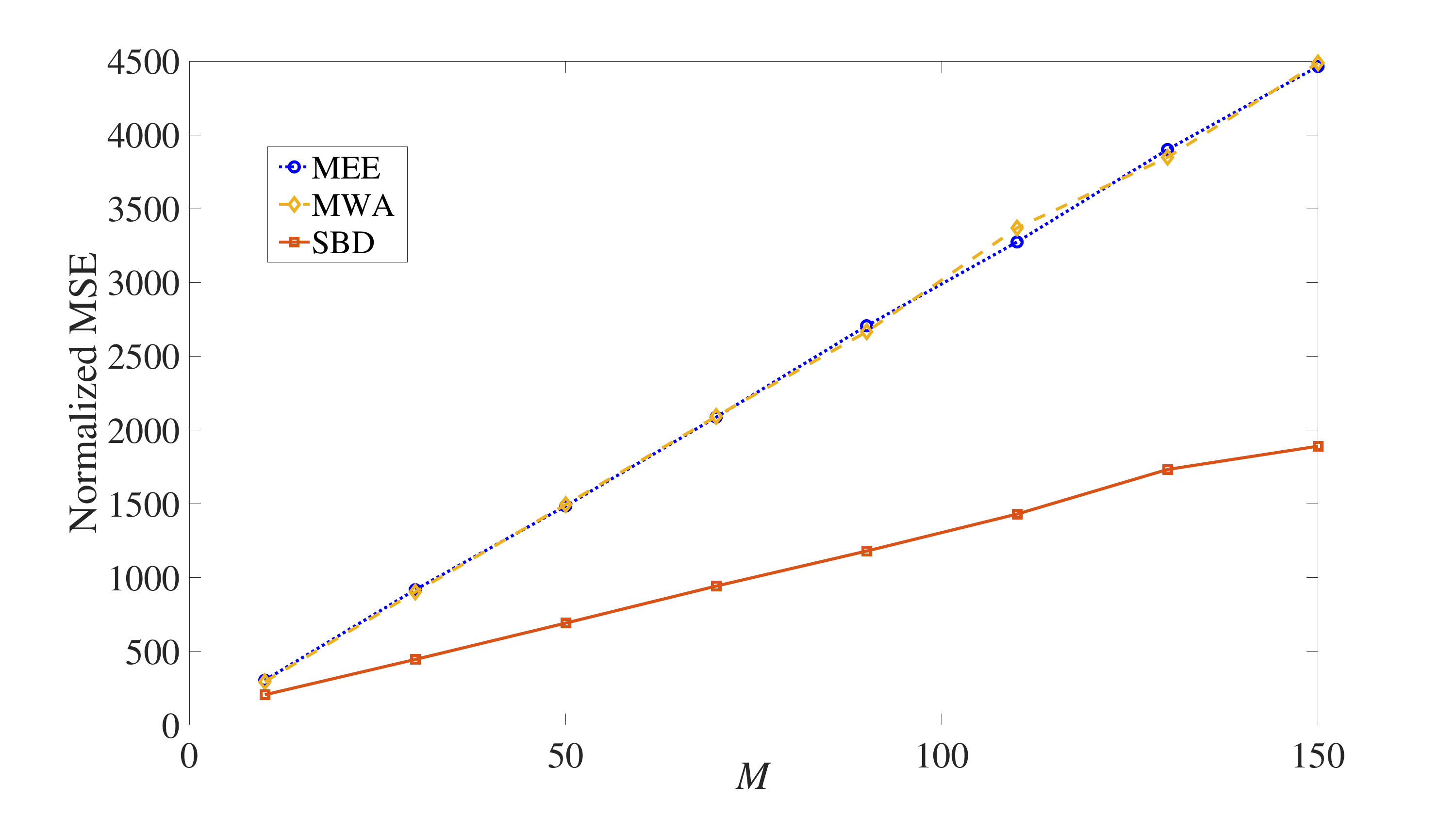}
	\caption{Normalized MSE of the SBD policy and the MEE and MWA policies against the number of sensor nodes $M$.}
	\label{fig_error_M}
\end{figure}

From the above results in Figs.~\ref{fig_error_b} and \ref{fig_energy_b}, it can be found that the parameters of $b$ and $k_{\mathrm{RP}}$ have an impact on both the normalized MSE and the energy cost. In order to compare the performance of the SBD policy with the MEE and MWA policies, we derived the optimal set of parameters $\{b^{*}, k_{\mathrm{RP}}^{*}\}$ of the SBD policy for different number of sensor nodes, which gives the smallest normalized MSE on condition that its energy cost is almost the same as the MEE and MWA policies. Fig.~\ref{fig_error_M} shows the normalized MSE of the SBD policy that employs the derived parameters and the MEE and MWA policies against the number of sensor nodes $M$. The results of the MEE and MWA policies coincide with each other since they behave similarly in this uncorrelated setting. Furthermore, we can see that the SBD policy shows a much smaller normalized MSE than the MEE and MWA policies. This is because the SBD policy manages to regularly collect observations with the large instantaneous error, while the MEE and MWA policies only aim to reduce the expected MSE over a long duration. These results confirm that the gain of exploiting instantaneous observations exceeds the cost of packet collisions, i.e., the effectiveness of the SBD policy for the uncorrelated case under the same constraint on the energy and radio resources. 

\subsection{Results for the correlated case}
\begin{figure}[t]
	\centering
	\includegraphics[width=0.5\textwidth]{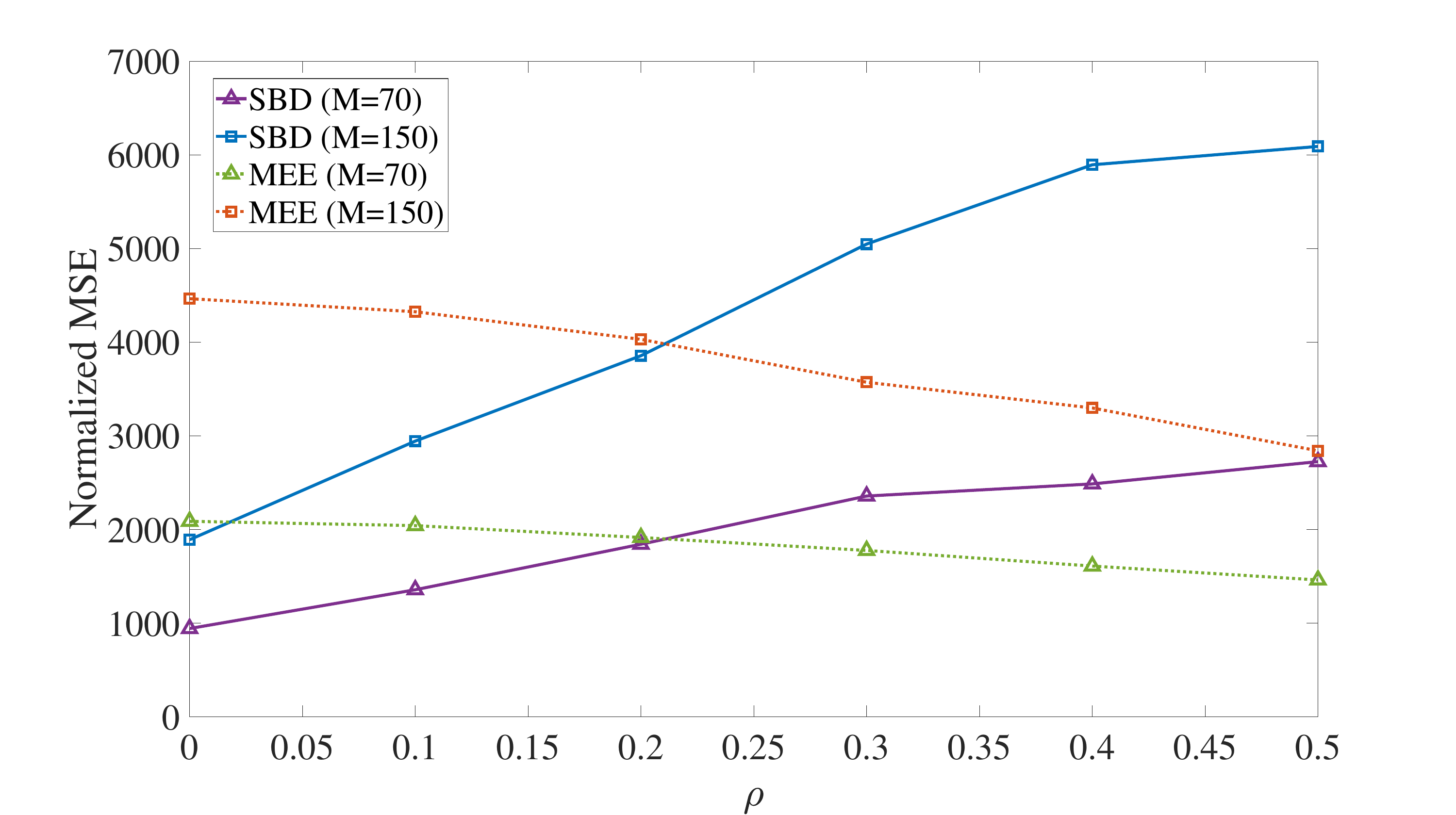}
	\caption{Normalized MSE of the SBD policy and the MEE policy against the degree of sensors correlation $\rho$.}
	\label{fig_error_rho}
\end{figure}

In this subsection, we investigate the impact of the correlation between sensor observations on the accuracy of the estimation. Fig.~\ref{fig_error_rho} shows the normalized MSE of the SBD policy and the MEE policy against the degree of correlation $\rho$ for the number of sensor nodes $M$ = 70 and $M$ = 150. Here, we only employ the MEE policy as an oblivious policy, since its superiority to the MWA policy for the correlated case has been shown in \cite{Ramakanth_Infocom}. The parameters of the SBD policy were derived based on the same approach as in the uncorrelated case. From Fig.~\ref{fig_error_rho}, we can see that the normalized MSE decreases as $\rho$ increases for the MEE policy. This is because the MEE policy collects data from a single sensor node at every scheduling instance without any collision, and its observation can be exploited to update the estimation of the other sensor nodes, especially when the degree of correlation is high. On the other hand, Fig.~\ref{fig_error_rho} shows that the normalized MSE of the SBD policy increases as $\rho$ increases. This is due to the increasing number of sensor nodes that simultaneously identify the necessity of transmissions for higher degree of correlation: when a single node observes the large deviation from its last reported observation, the other nodes are also likely to 
have similar observations. This increases the negative impact of collisions, causing a larger normalized MSE. However, we can see an improvement in terms of normalized MSE by employing the SBD policy for $\rho \le 0.2$.   

\section{Conclusions}
In this paper, we investigated how to achieve both low-power operations of sensor nodes and accurate state estimation using the Kalman filter for IoT monitoring employing wireless sensor networks under radio resource constraint. The conventional studies do not take into account energy consumption, therefore, we employed a wake-up receiver and wake-up signaling to reduce the standby energy of each sensor node. We designed a decentralized SBD policy including a wake-up signaling and random access, in which each sensor node autonomously transmits data employing a random access protocol after identifying its necessity to transmit data. By computer simulations, we showed that the SBD policy improves the accuracy of estimation in comparison to conventional oblivious policies on condition that the radio resource and energy cost are almost the same for the uncorrelated setting of sensor observations. We further clarified the degree of correlation in which the SBD policy shows superiority to a conventional oblivious policy.  

Our future work includes the design of a more elaborate random access scheme for a base station to retrieve informative data to improve the accuracy of the estimation.

\bibliographystyle{IEEEtran}
\bibliography{IEEEabrv,Ref}

\end{document}